\documentclass{iau-JDSS}
\usepackage{graphicx}

\title[Molecules and dust in evolved stars] 
{Spitzer observations of molecules and dust in evolved stars in nearby galaxies}

\author[M. Matsuura]   
{Mikako Matsuura%
}

\affiliation{Department of Physics and Astronomy, University College London, 
             Gower Street, London WC1E 6BT, United Kingdom; 
 email: mikako@star.ucl.ac.uk 
             }

\pubyear{2009}
\volume{Volume 15}  
\pagerange{119--126}
\date{?? and in revised form ??}
\jname{Highlights of Astronomy, Volume 15}
\editors{Ian F Corbett, ed.}
\begin{document}

\maketitle



\firstsection 

{\bf 1. Introduction}

Molecules and dust are formed in and around the asymptotic  giant branch (AGB) stars and supernovae (SNe), and are ejected
into the interstellar medium (ISM) through the stellar wind.
The dust and gas contain elements newly synthesised in stars, thus, dying stars play an important role on chemical
enrichment of the ISM of galaxies. However, quantitative analysis of molecules and dust in these stars had been difficult beyond our Galaxy.
The high sensitivity instruments on board the Spitzer Space Telescope  (SST; \cite[Werner et al. 2004]{Werner04}) 
have enabled us to study dust and molecules in these stars in nearby galaxies.
Nearby galaxies have wide range of metallicities, thus the impact of the metallicity on dust and gas production can be studied.
This study will be useful for chemical evolution of galaxies from low to high metallicities.

{\bf 2. Observations and results}

We present 5--35\,$\mu$m spectra, obtained by the infrared spectrometer (IRS) on board  the SST.
Targets are mainly asymptotic giant branch (AGB) stars located in five nearby galaxies within 200\,kpc.
The main objectives are 1) to exploit the dust property, molecular abundance, and mass loss of AGB stars
for a wide range of metallicities of the host galaxies, and 2) to investigate the contribution of mass loss from AGB stars on the galactic chemical evolution.
The metallicities of these galaxies range from half of the solar metallicity (the Large Magellanic Cloud) to 5\,\% of the solar metallicity (Sculptor dwarf spheroidal galaxy).
We found the following results.
\begin{itemize}
\item Carbon-rich AGB stars have abundant C$_2$H$_2$ molecules in low metallicity galaxies \cite[(Matsuura et al. 2005, 2006, 2007; Sloan et al. 2009)]{Matsuura06, Matsuura07, Sloan09}. 
At low metallicity, as the initial oxygen abundance is low,
after carbon elements are synthesised in stars, it leads to a high carbon-to-oxygen (C/O) abundance ratio
in carbon-rich AGB stars. A similar effect is found in amorphous carbon \cite[(Groenewegen et al. 2007)]{Groenewegen07}.

\item The SiC dust fraction is lower at lower metallicity in general, but this fraction suddenly increases at the end of the AGB phase \cite[(Gruendl et al. 2008)]{Gruendl08}.
The dust condensation sequence (SiC, and amorphous carbon/graphite) changes at low metallicity.

\item 
We found that the gas mass ejected from all AGB stars and SNe of gas are almost the same 
in the LMC  \cite[(Matsuura et al. 2009)]{Matsuura09}. AGB stars are one of the major dust sources in the LMC.
AGB dust at low metallicity tends to be more carbonaceous than oxides.
\end{itemize}

\end{document}